\documentclass[aps,pre,superscriptaddress,reprint]{revtex4-1}
\usepackage{amssymb,amsfonts,amsthm,bm}
\usepackage[ansinew]{inputenc}
\usepackage{amsmath}
\usepackage{graphicx}

\usepackage[pdfusetitle]{hyperref}

\newcommand{\Pg}{P_{\mathrm{c}}}
\newcommand{\dd}{\mathrm{d}}

\newcommand{\beq}{\begin{equation}}
\newcommand{\beqn}{\begin{eqnarray}}
\newcommand{\eeq}{\end{equation}}
\newcommand{\eeqn}{\end{eqnarray}}

\begin{document}

\title{Excimer formation of pyrene labeled lipids interpreted by means of coarse-grained molecular dynamics simulations}
\author{P. Ayoub, F. Thalmann$^{\ast}$}
\address{Institut Charles Sadron, Universit\'{e} de Strasbourg, CNRS UPR~22, 23 rue du Loess, Strasbourg Cedex, F-67037, France.\\
$^{\ast }$fabrice.thalmann@ics-cnrs.unistra.fr
}

\begin{abstract}{}{}{}
  The excimer formation dynamics of pyrene-labeled molecules in lipid bilayers depends on molecular  motion over distances of the order of 1-2~nm. From the concentration dependence  of the excimer photoemission curve, it is possible to derive a value for the lipid self-diffusion coefficient. This technique has been intensively used in the past twenty years, leading to rather large numerical values for self-diffusion compared with other approaches based on fluorescent probes tracking. In most cases, the interpretation of the experimental data rely on models for diffusion limited 2d reaction rates, or comparison with 2d lattice random walks. Our approach uses realistic molecular dynamics trajectories to reinterpret these experiments. Based on a well established coarse-grained model for lipid MD simulations (Martini), we show how to relate simulation results to experimental data on excimer formation. Our procedure is quite general and is applicable to all diffusion-limited kinetic processes. Key to our approach is the determination of the acceleration factor of lipid coarse-grained numerical models compared to reality. We find a significant reduction of the diffusion coefficient values, in particular when interleaflet association is taken into account. Our work does not point to deviation from a diffusion-limited mechanism but indicates that the excimer formation across bilayer leaflets could be hindered.
\end{abstract}


\maketitle

\markboth{Ayoub et al.}{Pyrene excimer formation dynamics}

\section{Introduction}
\label{sec:introduction}

Most lipid self-diffusion determination methods rely on long range diffusion (Fluorescence Recovery After Photobleaching FRAP, Fluorescence Correlation Spectroscopy FCS, Pulse Field Gradient NMR). On the other edge, inelastic neutron scattering probes the lipid dynamics on very short time and length scales. The pyrene excimer formation approach, initiated long ago~\cite{1974_Galla_Sackmann}, is unique as far its characteristic time  (100~ns) and length (1-2~nm) scales are concerned. It opens a window onto the nanometric scale in lipid membrane organization, currently subject to many supposed phenomena (rafts, static or dynamic nanodomains). How much is lipid diffusion dynamics on these scale regular or anomalous is a topic of great interest.

So far, numerical simulations are promising investigation tools to answer these questions. On the other hand, pyrene excimer formation experiments have been mostly interpreted in terms of 2d lattice random walks properties, which cannot describe faithfully the correlated motion of lipid molecules at this scale. There is little doubt that an analysis based on realistic molecular motion should outperform any approach based on point-like particles jumping over fixed discrete lattices.

In this article, we attempt to refine the past analysis of pyrene-lipid analogues diffusion by using coarse-grained (CG) molecular dynamics (MD) based on the celebrated Martini force-field~\cite{2007_Marrink_deVries}. Introducing a novel statistical approach that relates molecular dynamics trajectories to diffusion limited association kinetics among lipid chain subgroups, we obtain a numerical analogue to the excimer/monomer emission ratio as a function of the probe concentration.  This allows us to obtain the experimental diffusion coefficient by matching the short time excimer formation rate to the long time diffusion displacement.

We find diffusion coefficients values significantly smaller than previously admitted. This provides an important clue as far as reducing the spread among all experimental values currently found in the literature.

Our paper is organized as follows. Section~\ref{sec:excimerFormation} introduces the principle of excimer formation dynamics. Section~\ref{sec:Theory} explains how one can relate MD trajectories to experimentally relevant data. In Section~\ref{sec:NumericalResults} we analyze  data published by Vauhkonen et al., and Sassaroli et al.~\cite{1990_Vauhkonen_Eisinger, 1990_Sassaroli_Eisinger}. Results are discussed in Section~\ref{sec:Discussion} followed by Conclusion.

\section{Excimer formation dynamics}
\label{sec:excimerFormation}

Excimers are complexes formed by two identical molecules, one being in an excited state, the other in its ground state. The fluorescence emission spectrum of the isolated molecules, commonly referred as monomers, markedly differs from the emission spectrum of the excimers, or dimers. This makes it possible to measure optically the fraction of fluorescent probes ending up into an excited dimer (excimer) state. 

Excimer formation results from a diffusion limited mechanism of bimolecular reaction kinetics, and the formation rate is related to the self-diffusion coefficient of the fluorescent probes~\cite{1975_Birks}. The probability that an excited monomer ends up forming an excimer is the outcome of a competitive process involving bimolecular collision rate on the one hand, and spontaneous de-excitation rate (inverse fluorescence lifetime) on the other hand. 

Pyrene molecules form very stable excited dimer states. They can be incorporated, as pyrenyl groups, into phospholipid analogues and inserted in lipid membranes. These fluorophores are characterized by a longer than usual lifetime, ranging from 100 to 200~ns. The pyrene excimer formation is therefore reporting on the dynamical moves of lipid molecules on this time scale, corresponding to spatial displacements of the order of~1 to 2~nm in the bilayer. In this respect, this dynamics probes local molecular motions at a nanometric scale. The use of pyrene derivatives for the purpose of studying lipid membrane dynamics was reviewed by Somerharju~\cite{2002_Somerharju}. A larger class of diffusion-limited excitation or deexcitation processes in membranes is covered in a review by Melo and Martins~\cite{2006_Melo_Martins}.

The determination of lipid diffusion coefficients $D$ using pyrene derivatives is quite indirect. Fluorimetric techniques provides a ``titration curve'' of the monomer and excimer normalized emission intensities $J_M(x),J_E(x)$ as functions of the pyrene molar fraction $x$ and  monomer fluorescence lifetime $\tau_M$. Statistical models of molecular diffusion must be used to relate the diffusion constant to the excimer formation rate. So far, approaches based on cubic and hexagonal lattice random walks, and Smoluchovsky-Naqvi theoretical models have been used to interpret the experimental data~\cite{1974_Galla_Sackmann, 1990_Vauhkonen_Eisinger, 1990_Sassaroli_Eisinger,  1979_Galla_Sackmann, 1996_Martins_Melo, 2001_Martins_Melo, 2001_Novikov_Visser}. The approaches based on lattice random walks tend to give larger diffusion coefficient values than the ones obtained, \textit{e.g.}, by tracking fluorescent lipid probes over larger scales, such as in FRAP or FCS experiments (see Discussion). 

Current theoretical approaches are limited to ideal random walks or Wiener diffusion processes and they also involves approximations~\cite{1974_Naqvi,1974_Naqvi_Chapman,1983_Torney_McConnell,1989_Szabo}. Anomalous Brownian displacements of the pyrene groups, membrane heterogeneous nanodomains could both invalidate the current numerical estimates of lipid diffusion coefficients. Conversely, a robust microscopic description  of pyrenyl derivatives excimer formation would certainly help to improve the description of short range dynamical properties of lipid membranes.

In this work, we propose to use a well-known coarse-grained model for lipid bilayers, the Martini force field, and to analyze the excimer fluorimetric data based on realistic molecular dynamics (MD) trajectories.

\section{Theory}
\label{sec:Theory}

\subsection{First-passage reaction kinetics}

Diffusion-limited bimolecular reactions represent a class of physical or chemical kinetics in which reagents react at first encounter with a yield equal to, or close to unity. This limit means that there is no activation energy barrier, nor orientation barrier opposing the formation of the product~\cite{2010_Zhou}. 

The reaction rate of a diffusion limited reaction depends exclusively on the diffusion dynamics of the two species and on the geometrical shape of the reaction region. The most natural treatment, initiated by Smoluchovsky, considers two spheres following ideal Brownian trajectories and reacting as soon as their relative distance drops below a critical value: the \textit{capture radius} $\rho_c$. Naqvi derived a Smoluchowsky rate in the two dimensional case~\cite{1974_Naqvi}. Both approaches predict a time-dependent reaction rate and are restricted to ordinary (Wiener) diffusion processes. They are mean-field in nature as relative spatial correlations of the reactants are neglected~\cite{1983_Torney_McConnell}.

The experimental results reported in~\cite{1990_Sassaroli_Eisinger} were obtained with a pyrenyl group attached to a 10 atoms long alkyl chain. This group is mobile with respect to its parent molecule center of mass, and its Brownian displacement is affected by the intramolecular degrees of freedom on short time-scales. Its motion cannot be represented accurately by a Wiener process, as the associated \textit{mean-squared displacement} is not just a linear function of time and includes intramolecular Rouse dynamics. The consequences of such deviations with respect to the simplest Brownian model description, on the collisional dynamics properties of the pyrenyl groups must be assessed.  

The \textit{Martini} force-field treats each phospholipid molecule as an assembly of 10 to 14 beads, according to a 4~atoms to 1~bead correspondence. This level of coarse-graining preserves to a certain extent the intramolecular conformational degrees of freedom of the original atomistic model. With access to submolecular details, one can attempt to improve the collisional dynamics description compared with lattice or continuous models reducing lipids to point-like objects. This is the \textit{rationale} behind our treatment of pyrene derivative excimer formation dynamics. 

On the other hand, CG models are characterized by a nonphysical accelerated time scale, so that they cannot predict transport properties without a reference element (experiment or all-atoms simulation). The reference element, in our case, is the experimental monomer/excimer fluorescence ratio.

\subsection{Neutrality of fluorescent lipid probes substitution}

Fluorescence is an extremely useful and versatile technique. Its use in the field of membrane studies includes imaging, life-time,  anisotropy depolarization and resonant energy transfer studies. In most cases, the amount of necessary fluorophores is small, as fluorescence detection can be very sensitive. A common assumption of all these approaches is that the fluorescent labeled molecules, which often are made as similar to phospholipid molecules as possible, modifies only marginally the structure and the dynamics of the host membrane. For instance, the diffusion coefficient obtained in a FRAP experiment, is the one associated with the fluorescent probes. The identity between probe and normal lipid motions remains an assumption. 

The pyrene excimer formation method requires relatively high amounts of pyrenyl derivatives, up to 10\% molar ratio. It is assumed that the motion of the pyrenyl-grafted lipids is identical to the regular lipid compounds, and that the bilayer structure is not modified significantly. We therefore adopt this view and treat pyrene-labeled lipids as if they were ordinary phospholipid molecules. This is both a simplifying assumption and a limit of the present approach. 

We do not have at our disposal a coarse-grained model of pyrene-labeled lipid molecule. The poly\-aromatic pyrenyl group is clearly bulky compared with standard alkyl chains, and the effective van-der-Waals interaction parameters should be increased to reflect its higher polarizability. It is difficult to estimate these parameters at coarse-grained level, and guesswork is unlikely to provide an acceptable set of interactions. Another difficulty arises when it comes to model the interaction between a photoexcited pyrene group and one in its ground state. It is believed that attractive interactions arise from resonant energy states, leading to stronger and longer range ( $r^{-3}$ with separation $r$) attractive forces, eventually leading to the complexed dimer state~\cite{1975_Birks, 1964_Stephen}. In a realistic model, those modified interactions should be estimated and taken into account. 

Even if a CG model of pyrenyl derivatives existed, simulating a diluted mixture of probes and repeated sequences of excitation $\to$ diffusion $\to$ excimer formation would require lengthy and multiple runs, to obtain just a rough statistics of the events. Our opinion is that such an intense computational effort would only make sense if a trusted and robust CG model of pyrenyl compounds was first established, including the modified interactions of the photoexcited pyrene group.  

We remain faithful to the implicit assumption that fluorescent probes behave the same as the major lipid components. Ordinary MD simulation runs are performed and trajectories acquired. The equilibrated trajectories are then analyzed \textit{a posteriori}, as if a few of the lipid beads were actually pyrenyl moieties. Our strategy is therefore  to \textit{rerun} the trajectory (without recomputing  the forces) with a relabeling of some beads \textit{a posteriori}. In practice, one can reduce the determination of the pyrene excimer formation assays to regular dynamical time dependent correlation functions of the ordinary lipid molecules. The next subsection introduces such dynamical quantities.

\subsection{Survival probabilities}

\begin{center}
\begin{figure}
\resizebox{0.42\textwidth}{!}{\includegraphics{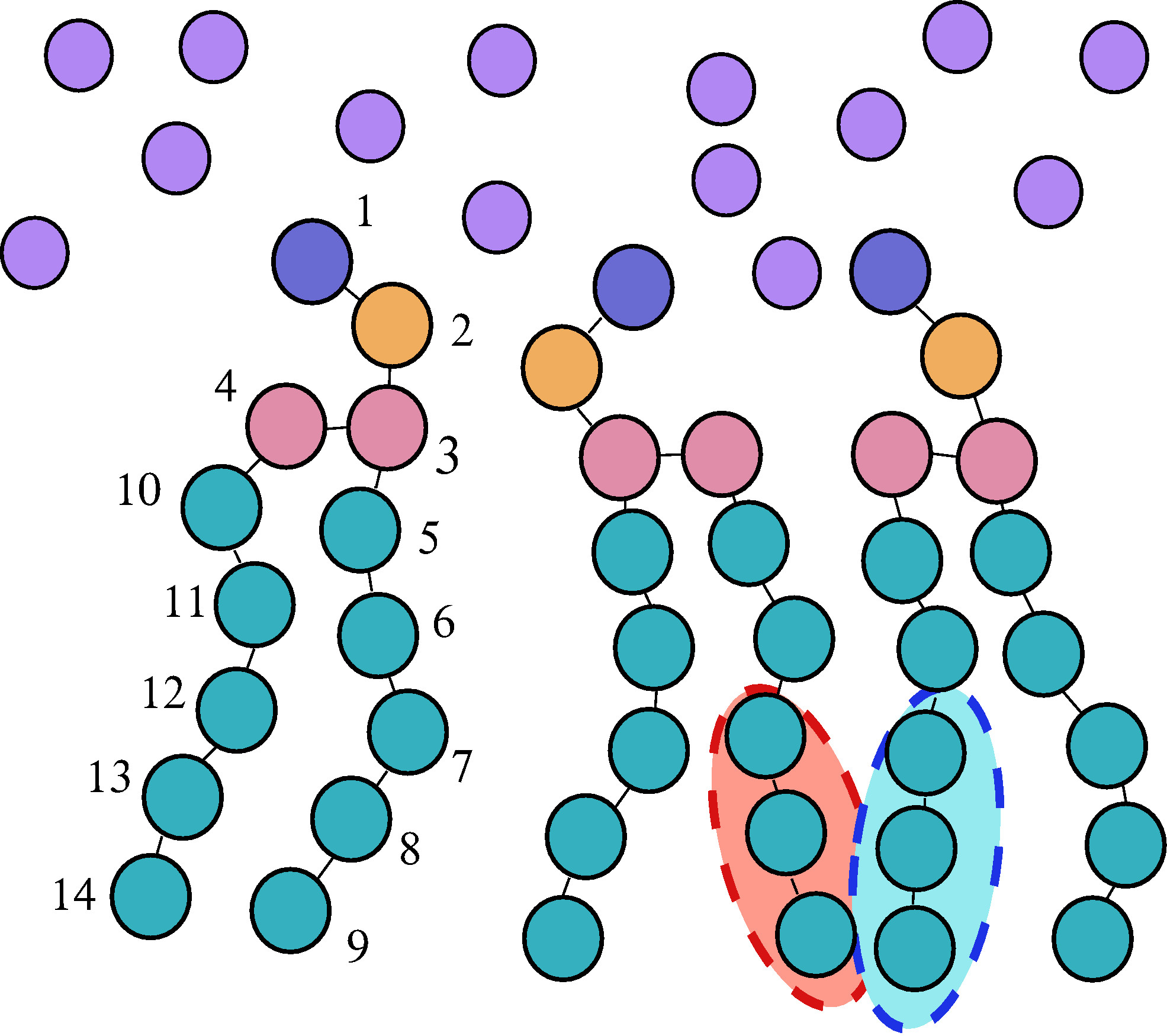}}
\caption{Example of coarse-grained lipid molecule (DOPC) with bead labels, and two terminal subgroups during a ``collision event''.}
\label{fig:CG-lipid}
\end{figure}
\end{center}

\begin{center}
\begin{figure}
  \resizebox{0.42\textwidth}{!}{\includegraphics{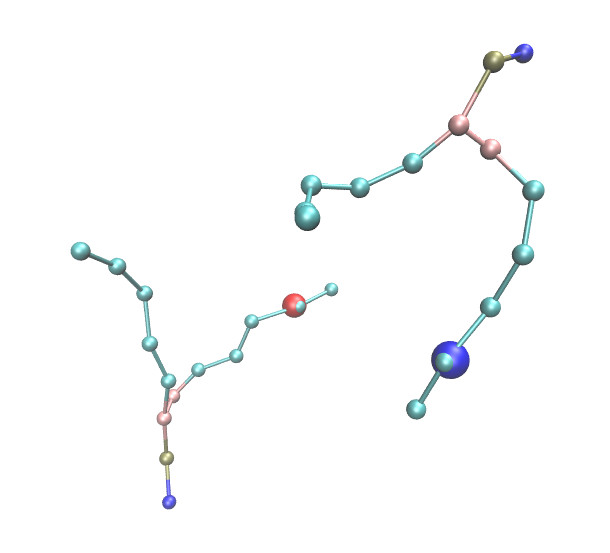}}
  \caption{Detail of two DOPC molecules, each in one leaflet, with terminal subgroups superimposed on top of them.}
\label{fig:detail2Lipids}
\end{figure}
\end{center}

\begin{center}
\begin{figure}
  \resizebox{0.5\textwidth}{!}{\includegraphics{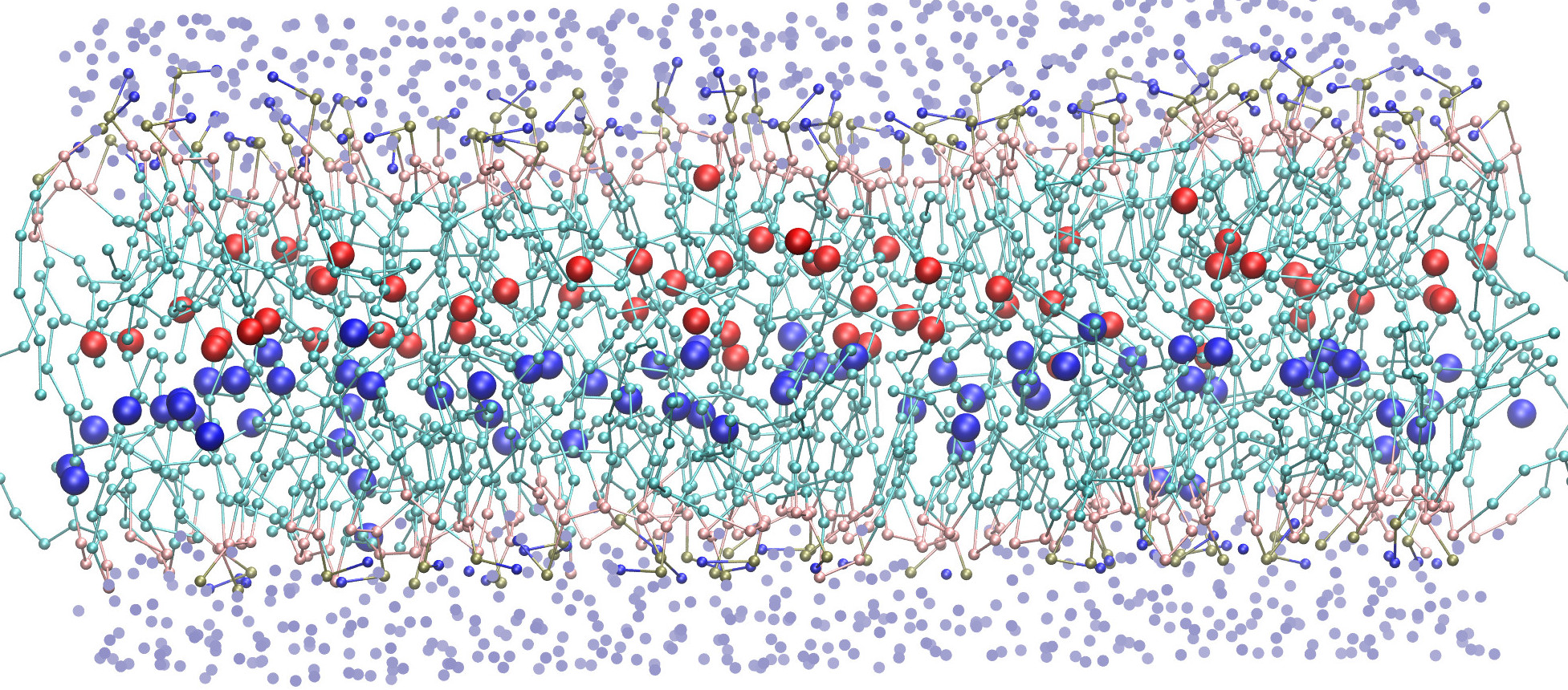}}
  \caption{Side view of a DOPC bilayer, with superimposed terminal subgroups from the upper and lower leaflets (using two different colors depending on the leaflet to which the molecule belongs).}
\label{fig:globalBilayer}
\end{figure}
\end{center}

A CG model of DOPC is represented on Fig.~\ref{fig:CG-lipid} and Fig.~\ref{fig:detail2Lipids}. We propose to assign to the pyrene group a set of three beads numbered from 12 to 14, hereafter referred as the \textit{the fluorescent group}, or \textit{terminal subgroup} (TS). A similar procedure defines the POPC terminal subgroup (not shown).  We assume that the excimer association takes place as soon as the relative distance between two terminal subgroups becomes less than a critical radius, which we refer as the  \textit{capture radius} $\rho_c$, by reference to the Smoluchowsky description of first passage reaction kinetics. Fig.~\ref{fig:globalBilayer} shows a bilayer snapshot with its associated terminal subgroups. 

Let us now consider a simulation box containing a total of $N_t$ lipids, among which $N_p$ will be considered \textit{a posteriori} to be fluorescently labeled, in each leaflet. We denote $L$ the side of a squared simulation box in the $x$, $y$ directions, with periodic boundary conditions (pbc). We introduce the survival probability $P_s(\rho_c,N_p,N_t,L; t)$ as the probability that no excited fluorescent group have come into contact with any of the $N_p-1$ remaining groups \textit{in the same leaflet} during a time interval $t$.

$P_s$ is the outcome of an averaging procedure over both initial conditions (the choice of the fluorescent groups and their actual spatial distribution) and subsequent trajectories (the Brownian displacements of these fluorescent groups). The negative time derivative $-\dd P_s/\dd t$ represents the probability of ``capture'' per unit of time, or first passage time distribution. By \textit{capture} is meant that one of fluorescent group in its ground state penetrates into a sphere of radius $\rho_c$ centered around the excited group for the first time during the interval $[t,t+\dd t]$. 

We define in a very similar way $P_o(\rho_c, N_p,N_t,L; t)$ the probability that any pair of fluorescent groups taken from randomly selected molecules \textit{in two opposite leaflets}, have not come closer than a distance  $\rho_c$ during a time $t$. The collision dynamics controlling the formation of excimers is order of magnitude faster than the phospholipid flip-flop reversal times so that it makes sense to distinguish between the two subpopulations of lipids, and $P_o$ effectively controls the influence of interleaflet monomer quenching.   

The survival probabilities $P_s,P_o$ can be efficiently sampled from regular MD trajectories. These dynamical quantities are well defined and related to the molecular displacements properties of the corresponding groups of beads. As we deal with realistic trajectories, we can address situations with transient dynamics, or non Gaussian self-intermediate scattering functions, and possibly improve upon most basic models of molecular diffusion. It remains, however, to relate these probabilities to the excimer/monomer fluorescence ratios which are experimentally obtained.

\subsection{Size-scaling and independent pairs assumptions}

Let us consider a lipid bilayer containing a total of $N_t$ lipids in its upper leaflet, including a number $N_p$ of fluorescent probes. For simplicity, we assume that both leaflets share the same composition, making the bilayer symmetric. The size of the bilayer is supposed to be unaltered by the presence of the probes, so that the size $L$ of the system is directly related to the number of lipids $N_t$ by means of the area per lipid $a_0$ ( $N_t=L^2a_0$).  

Under normal illumination conditions, the fraction of time spent in the excited state is tiny, and at any given time the chances of finding two excited monomers in the same portion of membrane is statistically very low. One can therefore consider that one fluorophore at most is excited at a time, and that $N_p-1$ monomers are available to combine as an excimer. 

If the motions of the pyrene monomers are decorrelated, the survival probability $P_s(\rho_c, N_p,N_t; t)$ can be reduced to $P_s(\rho_c, 2,N_t; t)^{N_p-1}$. This is because $N_p-1$ different pairs compete independently, and that the effective collision time is \textit{the minimum} of $N_p-1$ independent first passage events. Note that the interaction between pyrene monomers is only possible when at least two of them are situated at the immediate vicinity of an excited probe. The outcome of such a situation involving a compact triplet of probes is difficult to ascertain, either enhancing, or disfavoring the formation of a stable excimer. We therefore refer to \textit{independent pairs assumption} the possibility of factorizing the survival probability when the concentration $x$ of probes is low enough.  

When interleaflet interaction is allowed, the survival probability in the independent pairs assumption framework reads:
\beq
\Pg(t) = P_s(\rho_c, 2,N_t,L; t)^{N_p-1}P_o(\rho_c, 1,N_t,L; t)^{N_p}.
\eeq

We now address the issue of the sample size dependence. It sounds quite natural to assume that the survival probability depends on size ($L$ or $N_t$) only throughout the intensive ratio $x=N_p/N_t$. If this holds, the survival probability for collisions between opposite leaflets, for instance,  should obey the following scaling relation:
\beqn
P_o(\rho_c, N_p,N_t; t) &=& P_o(\rho_c, \lambda^2 N_p,\lambda^2 N_t; t)\notag\\
&=& P_o(\rho_c, \lambda^2 x N_t,\lambda^2 N_t; t).
\eeqn
In the case of collisions between groups in the same leaflet, we observe that the excited pyrene group, \textit{i.e.} the target, must be removed from the list of moving groups participating to the excimer quenching statistics. This has consequences when $N_p$ is small, and the suggested size scaling form for the survival probability within a given leaflet must be slightly modified to account for the ``missing'' pyrene group.
\beqn
P_s(\rho_c, N_p,N_t; t) &=& P_s(\rho_c, 1+\lambda^2 (N_p-1),\lambda^2 N_t; t)\notag\nonumber\\
&=& P_s(\rho_c, 1+\lambda^2 (x N_t-1),\lambda^2 N_t; t).\notag\\
\label{eq:survivalProba}
\eeqn

We refer to these assumptions as the \textit{size scaling} assumptions. There are reasons to believe, though, that the size scaling assumptions do not hold uniformly. As a matter of fact, collision time distributions do depend explicitly on sample size in 2d Brownian reaction-diffusion models. For instance, the probability of finding a target in a domain of linear size $L$ involves $\ln(L)$ corrections with respect to the naive mean-field result. We claim however that these corrections are unlikely to change very much the short time dependence of the survival probability. As far as pyrene excimer formation is concerned, the contribution of long trajectories wandering on large distances $\sim L$ away from the capture radius are unlikely to contribute to the short time behavior of the survival probability, which turns out to dominate the excimer formation probability. However, they would give rise to significant contributions at longer times.

Combining \textit{independent pairs} and \textit{size scaling} assumptions, we can reduce the collision survival probability of an excited pyrenyl group to the survival probability of a single pair sitting in a bilayer of arbitrary size $\lambda L$: 
\beqn
\Pg(t) &=& P_s(\rho_c,2,\lambda^2 N_t; t)^{\lambda^2 x N_t-\lambda^2}\notag\\
& &  \;\;\times\; P_o(\rho_c,1,\lambda^2 N_t; t)^{\lambda^2 x N_t},\notag\\
&\simeq&  P_s(\rho_c,2,\lambda^2 N_t; t)^{\lambda^2 x N_t}\notag\\
& &  \;\;\times\; P_o(\rho_c,1,\lambda^2 N_t; t)^{\lambda^2 x N_t}.
\label{eq:x-Ps-dependence}
\eeqn
The resulting survival function $\Pg(t)$ is the global collision survival probability. 

What remains to be done is to use a scaling factor $\lambda$ corresponding to a tractable simulation scheme. In the present work, we use $\lambda^2 N_t = 256$ to sample $P_s$ and $P_o$. In addition, we notice that in experimentally relevant situations, $\lambda^2 N_t$ is significantly smaller than the actual number of lipids in the physical system of interest (\textit{e.g.} a liposome), and the $\lambda^2$ term in the exponent of $P_s$ can be safely neglected in eq.~(\ref{eq:x-Ps-dependence}).

\subsection{From survival probabilities to excimer fluorescence intensity}

The excimer formation probability is readily obtained from $\Pg(t)$. A monomer excited at $t=0$ forms an excimer during the time interval $[t,t+\dd t]$ with probability 
\beq
-\frac{\dd \Pg}{\dd t} \exp\left(-\frac{t}{\tau_M}\right), 
\eeq
which is the probability of colliding with a ground state monomer while still in the excited state, with $\tau_M$ standing for the monomer fluorescence lifetime in the ultra-dilute case ($x \to 0$). The excimer formation probability $J_E$ is therefore
\beqn
J_E &=& \int_0^{\infty} -\frac{\dd \Pg}{\dd t} \exp\left(-\frac{t}{\tau_M}\right) \dd t \notag\\
&=& 1-\int_0^{\infty} \Pg(t) \exp\left(-\frac{t}{\tau_M}\right) \frac{\dd t}{\tau_M}.
\label{eq:Survival2Titration}
\eeqn
$1-J_E$ is the Laplace transform of the survival probability. Naturally, the probability of returning to ground state from the excited monomer state is $J_M=1-J_E$. The  probe ratio $x$ dependence of the monomer/excimer fluorescence intensity curves $J_M(x),J_E(x)$  comes from the $x$ dependence of the survival probability $\Pg(t)$, such as expressed in eq.~(\ref{eq:x-Ps-dependence}).
Alternatively, the integrand
\beq
\frac{1}{\tau_M}\Pg(t)\exp\left(-\frac{t}{\tau_M}\right)
\eeq
corresponds to the time-resolved emission intensity of the monomer probe.

\subsection{Intermolecular formation rates}

In principle, the survival probability $\Pg$ must be extracted from the microscopic model for any arbitrary value of $x$, and the normalized emission intensities $J_E$ and $J_M$ become dependent on the molecular ratio $x$. The variation of the fluorescence intensities therefore reports on the concentration of monomeric (or dimeric) probes, and behaves as in a standard titration experiment, except that for practical reasons $x$ cannot be changed during the course of the experiment.

One peculiar value $x=x^{\ddagger}$ makes the excimer formation probability equal to $1/2$: 
\beq
J_M(x^{\ddagger}) = J_E(x^{\ddagger}) = 1/2.
\eeq
This value turns out to plays a key role in the interpretation of the experimental titration curves $J_E(x),J_M(x)$ by Vauhkonen et al.~\cite{1990_Vauhkonen_Eisinger} .

Let us assume first that the survival probability $\Pg(t)$ is exponentially decaying as $\exp(-t/\tau_c(x))$. In this case, $x^{\ddagger}$ is nothing but the concentration for which the collision time $\tau_c(x^{\ddagger})$ equals the monomer spontaneous decay time $\tau_M$,
\beq
\tau_c(x^{\ddagger})=\tau_M.
\eeq

It is common to treat the excimer formation dynamics as an ordinary \textit{bimolecular kinetic process} $M+M^{\star} \to (MM)^{\star}$ with formation rate $K(x)$, and to take the reverse dissociation as negligible. As the monomer is in excess, the situation corresponds to an ideal pseudo-first order reaction kinetics, consistent with an exponential decay of the excited monomer population.  The decrease rate of the excited probes reads $\dd [M^{\star}] = -(K(x)+\tau_M^{-1})[M^{\star}]\dd t$, leading to the monomer emission intensity
\beq
\frac{1}{J_M(x)} = 1+\tau_M K(x).
\eeq
For a pure mean-field, constant rate bimolecular mechanism, $K(x)$ is expected to depend linearly on $x$, enabling us to rewrite the previous expression as
\beq
\frac{1}{J_M(x)} = 1+\frac{x}{x^{\ddagger}}.
\label{eq:from_constant_rate}
\eeq

The corresponding normalized emission intensity curves then reduce to simple rational functions.
\beq
J_M(x) = \frac{x^{\ddagger}}{x+x^{\ddagger}}\; ; \; J_E(x) = \frac{x}{x+x^{\ddagger}}.
\eeq
It can be shown that pseudo-first order excimer formation kinetics and exponential survival behavior are two equivalent assumptions. Conversely, any deviation from linear behavior of the inverse emission intensity $1/J_M(x)$ points to a \textit{non exponential survival behavior} of the excited monomeric probes, or a \textit{non constant rate mechanism}. Example of diffusion limited models leading to such time dependent rates are discussed, for instance, in~\cite{1983_Torney_McConnell}. Our analysis does not make any assumption regarding the kinetic rate coefficients, and the survival probability is sampled from MD trajectories.

\subsection{Multiple passage times}

The lack of detailed evidence for the molecular excimer formation mechanism makes it necessary to adopt a phenomenological and probabilistic view on the outcome of the collisional dynamics. Even though the excimer formation is expected to be diffusion dominated, there is nevertheless a chance that collisions does not lead to excimer state with perfect yield.

It is possible to introduce a distance dependent excimer formation rate, such as \textit{e.g.} the distance dependent energy transfer rate of a F\"{o}rster resonant pair. In this case, the excimer formation probability depends on the integrated time spent by the pair at close distance.

Denoting by $r(t)$ the relative distance of a single pair, the survival probability of an excited monomer associated with a trajectory $\{r(t)\}$ reads 
\beq
\exp\left( -q\int_0^t \chi\left[\frac{r(t')}{\rho_c}\right]\,\dd t' \right),
\label{eq:tunableRate}
\eeq
with $q$ denoting a tunable excimer formation rate, $\rho_c$ a characteristic capture radius, and $\chi$ a positive function decaying from~1 to~0.

The generalization of the survival probability is obtained by averaging over the initial positions and subsequent trajectories $r(t)$:
\begin{multline}
P_s(q,\rho_c, 2,\lambda^2 N_t;t) =\\
\left\langle\exp\left( -q\int_0^t \chi\left[\frac{r(t')}{\rho_c}\right]\,\dd t' \right)\right\rangle_{\{r(t)\}},
\label{eq:SurvivalFunctionSampling}
\end{multline}
with the same notation as in eq.~(\ref{eq:survivalProba}).   One notices that the first passage survival probability can be obtained as a limit case of a step function $\chi[v] = 1$ for $v\leq 1$ and $0$ otherwise, combined with $q\to \infty$. The survival probability for an arbitrary $x$ concentration follows from the size-scaling relation.

Functions $P_s,P_o$ defined as in eq.~(\ref{eq:SurvivalFunctionSampling}) can be conveniently sampled from molecular dynamics trajectories. This approach to multiple passage dynamics is simple to implement, and contains the first passage capture process as a limit case.

Low values of $q$ correspond to low excimer formation rate. A general feature, as we found, is that the resulting $P_s(t)$ becomes more and more exponential as $q$ decreases, with an average survival time increasing as expected. The resulting inverse probability $1/J_E$  is therefore expected to become more and more linear with $x$ as $q$ decreases, giving us a chance of estimating $q$. This turned out not to be the case in practice. 


\subsection{Estimating the capture radius}

There is arbitrariness when it comes to giving a numerical value to the reaction, or capture radius $\rho_c$. In the Martini model, the van der Waals radii of the coarse-grained beads are set to 0.47~nm. Closer approach between beads can only be associated to enthalpic repulsive interactions, which are not supposed to intervene in a diffusion-limited association process. 

\begin{center}
\begin{figure}
\resizebox{0.46\textwidth}{!}{\includegraphics{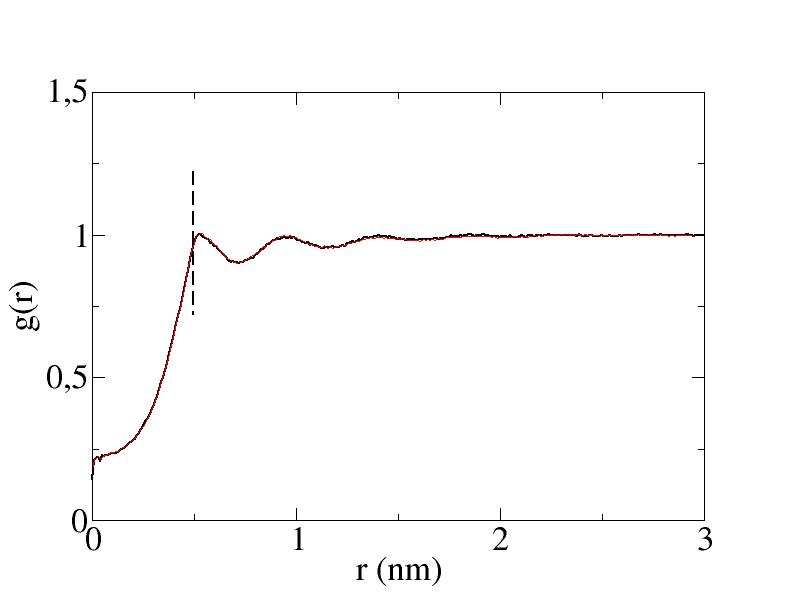}}
\caption{Pair distribution of terminal subgroups. The vertical dashed line indicates the horizontal separation $r=0.5$.}
\label{fig:pair-distribution-groups}
\end{figure}
\end{center}

On the other hand, in order to represent as well as possible  the motion of pyrene groups, we bunched together the last beads of each hydrophobic chain groups. The generalized coordinates representing the ``pyrene'' center of masses, are themselves not materialized as beads and their mutual interaction potential is softer, to some extent, than the one of the original beads. Fig.~\ref{fig:pair-distribution-groups} represents the 2d pair distribution function of group belonging to a given leaflet.

It can be seen that steric hindrance between pyrene center of masses becomes significant for values of $r$ smaller than 0.5~nm. We therefore take $\rho_c=0.5$~nm as our capture radius value.

\subsection{Real systems vs coarse-grained dynamics}

A first question arises, as to which extent coarse-grained dynamics is faithful to the trajectories of the real phospholipid molecules that it is intended to model. It is established that most large scale transport properties such as self-diffusion, transverse viscosity, etc are \textit{in fine} controlled by \textit{cis-trans} isomerization dynamics of the lipid alkyl chains~\cite{Cevc_Marsh_PhospholipidBilayers}. These degrees of freedom are obviously missing at the level of the Martini coarse-grained description, and substituted for by an effective ``Kremer-Grest'' dynamics of beads and springs polymer chains subject to van-der-Waals interaction. This accounts for the largest fraction of the difference seen between the coarse-grained and atomistic molecular kinetics. Martini simulation are sped up by an acceleration factor $f$ commonly taken of the order of~4. As a by-product, CG molecular dynamics cannot, alone, provides quantitative estimates of the lipid transport coefficients. 

The \textit{cis-trans} chain isomerization moves are temperature dependent, activated processes. As a result, the membrane fluidity strongly depends on temperature, showing  Arrhenius dependence in the absence of phase transition. The Martini
activation energy is lower than the experimental one. The factor $f$ is therefore likely to depend significantly on temperature. On the other hand, it is not necessary to match precisely the temperature of the CG simulation with the experimental system that one aims at reproducing, provided one restrict ourselves to the same structural phase. We may therefore rely on a kind of  \textit{time-temperature} superposition approximation to adjust to experimental data. Note that irrespective to changes occurring with lipid dynamics, the monomer fluorescence life-time is also a decreasing function of temperature, and must be provided as an external experimental input. 

We assume in our approach that the acceleration factor $f$ applies in a uniform way, from short chain reorganization time scales up to long range hydrodynamic limit. In other words, if one can determine $f$ based on the intermediate time regime associated with fluorescence life-time and collision induced excimer formation, one should be able to give a quantitative prediction of the physical lipid diffusion constant $D$ as $D=D_{\mathrm{MD}}/f$, where $D_{MD}$ represents the molecular dynamics diffusion coefficient obtained from the mean-squared displacement. This can be assimilated to a \textit{matching procedure}. 

Another difference between real molecules and CG models lies into the repartition of masses within lipids. The Martini model ascribes an identical mass (72~amu) to all beads, while in reality the phosphate headgroup concentrates more mass than the alkyl chains for comparable steric volumes. As dynamical properties depend on masses (unlike thermodynamics) this could cause differences between real and simulated trajectories. In the present work, we assume that such differences can be disregarded. 


\section{Numerical results and comparison with experiments}
\label{sec:NumericalResults}

\subsection{Diffusion properties of the terminal subgroups, and survival probabilities}

Survival probabilities were sampled by simulating patches of 512 lipids, respectively POPC at 293K and DOPC at 283K, using 4 trajectories of 150~ns (Martini time). Frames were recorded every 6~ps for the collision statistics, and averages over all the possible pairs of molecules were taken in order to sample $P_{s}(t,2,256)$ and $P_{o}(t,1,256)$. Trajectories were generated with Gromacs-4.6~\cite{2008_Hess_Lindahl} using a NVT Nose-Hoover thermostat scheme, and the Martini force-field \textit{v2} for the lipids.

\begin{center}
\begin{figure}
\resizebox{0.42\textwidth}{!}{\includegraphics{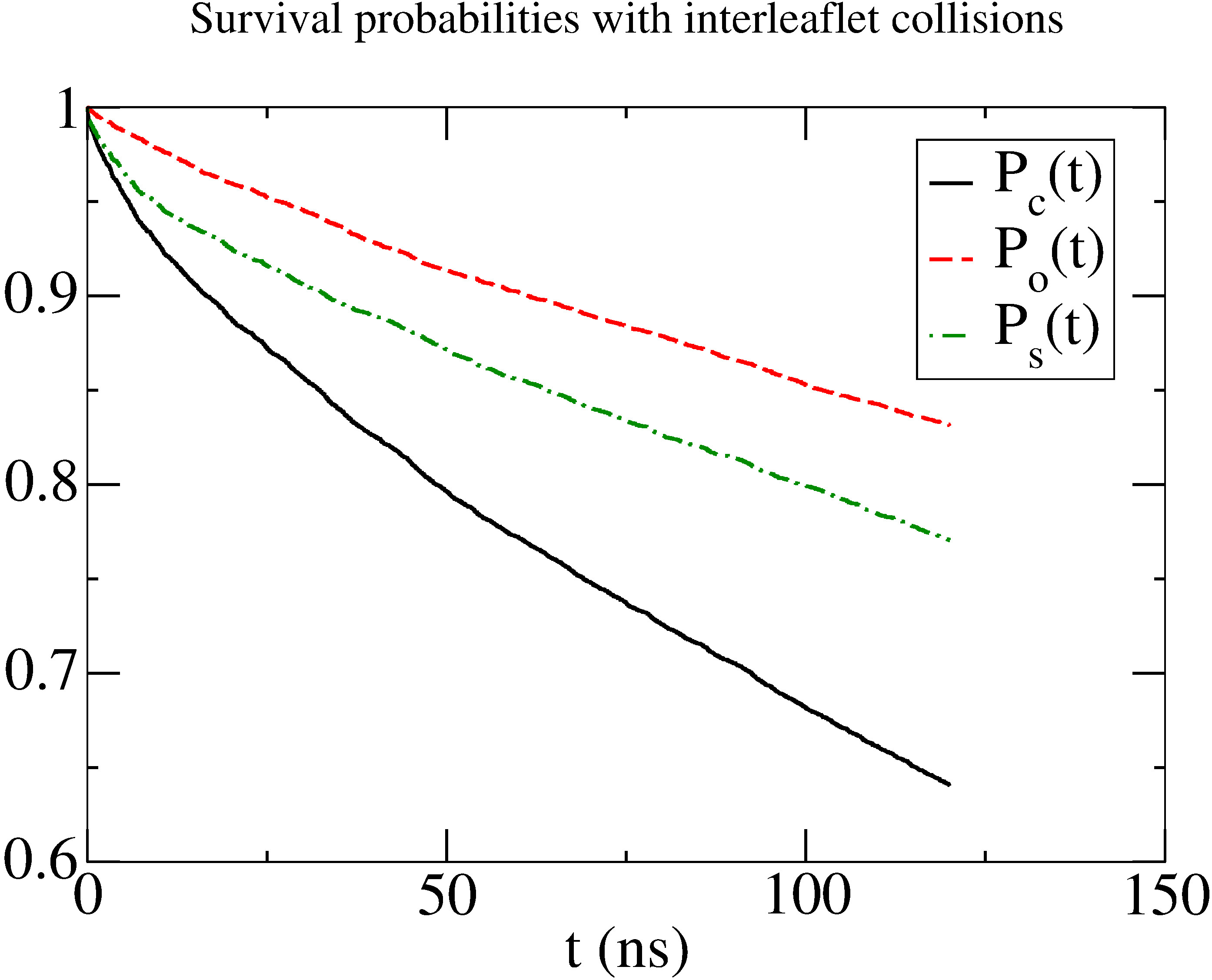}}
\caption{Survival probabilities restricted to same leaflet, opposite leaflet and product of both. Trajectories are from DOPC at 283~K, with a capture radius $\rho_c=0.5$~nm and frames taken every 6~ps of simulation time.}
\label{fig:productSurvival-DOPC}
\end{figure}
\end{center}

Fig.~\ref{fig:productSurvival-DOPC} shows the survival probability when collisions are restricted to molecules belonging to the same leaflet $P_s(2,256,t)$, to opposite leaflets $P_o(1,256,t)$, and then the product $P_c=P_oP_s$. The latter is expected to rule unrestricted  excimer formation. Introducing collisions between opposite leaflets increases significantly the survival decay rate.

\begin{center}
\begin{figure}
\resizebox{0.42\textwidth}{!}{\includegraphics{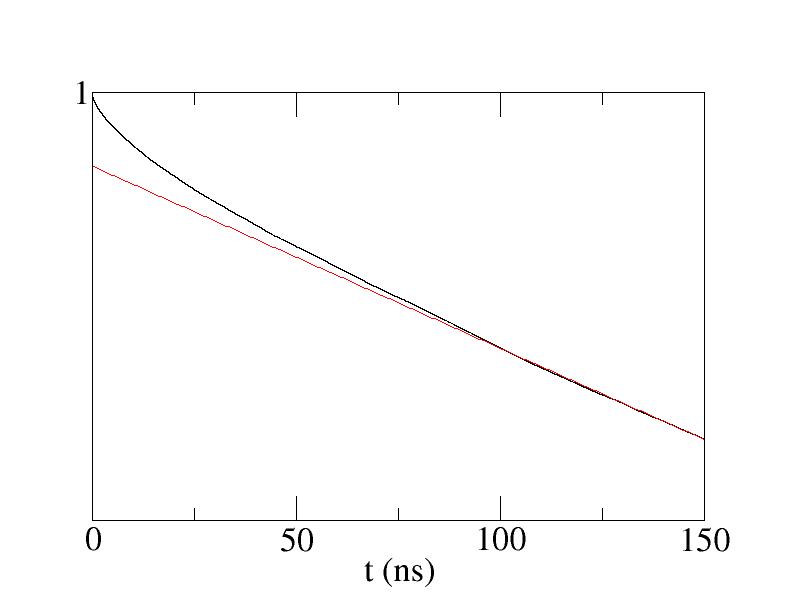}}
\caption{Survival probability $\Pg(t)$ when collisions are unrestricted (DOPC at 283~K). The vertical axis is logarithmic, and a straight line representing the asymptotic behavior $0.926\exp(-0.00196 t)$ ($t$ in ns) is also shown. The straight line goes from 0.926 ($t$=0) to 0.687 ($t$=150~ns), and the distance to upper curve represents a deviation from pure exponential behavior of $\Pg$. The curve represented above corresponds to a concentration $x=1/256$, and the non-exponential behavior at short times is reinforced as the concentration $x$ increases.  } 
\label{fig:logSurvival-DOPC}
\end{figure}
\end{center}

Fig.~\ref{fig:logSurvival-DOPC} shows the collisional survival probability $P_c(2,256,t)$ of the chain subgroups, that is central to our reaction kinetics modeling, on a semi-logarithmic plot.  As discussed in the previous section, such non exponential behavior can be related to deviation from the Stern-Volmer linear plot of the titration curves. %


\begin{center}
\begin{figure}
  \resizebox{0.42\textwidth}{!}{\includegraphics{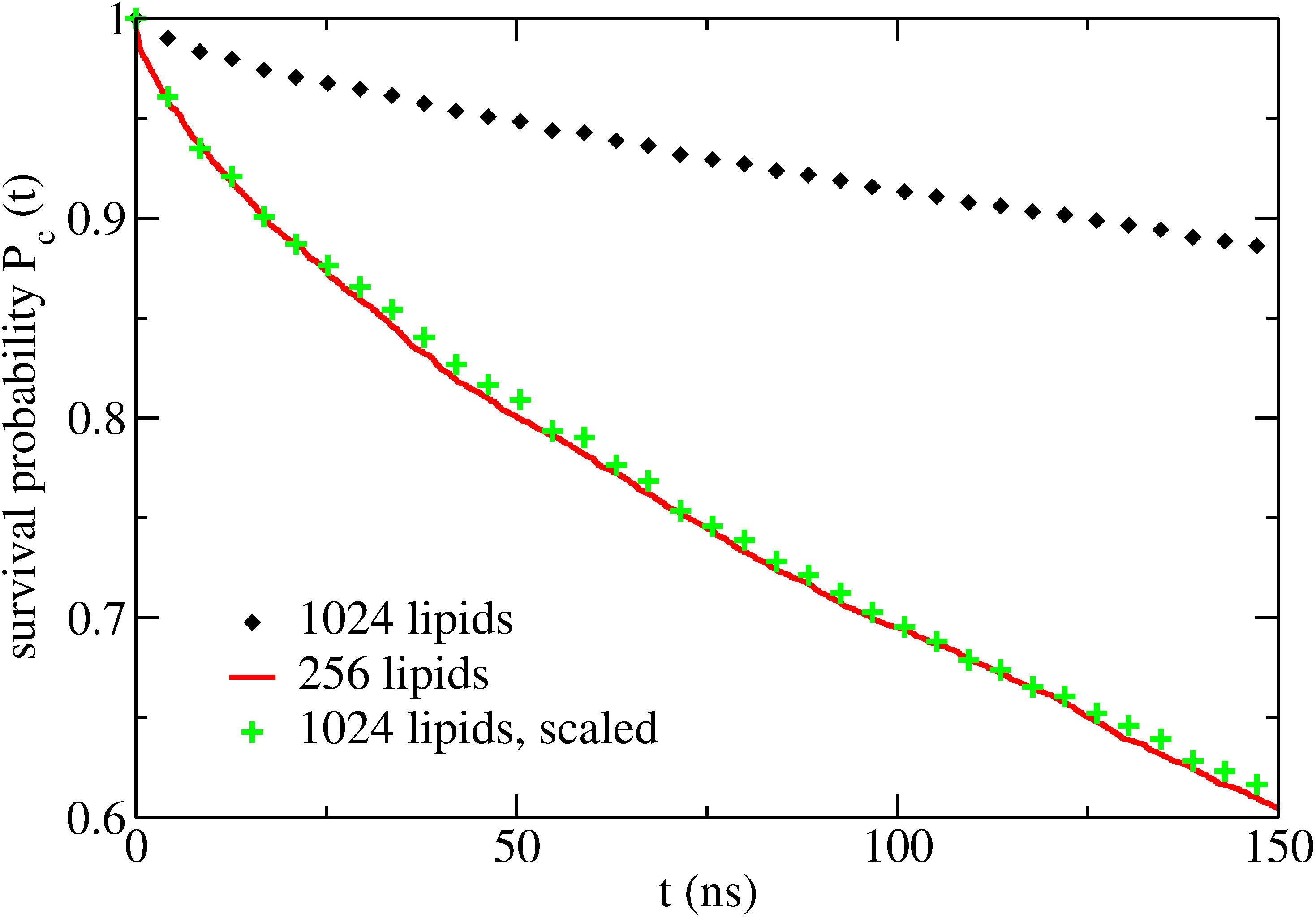}}\\
\caption{Size-scaling approximation for DOPC molecules, using 512 and 2048 lipids in the simulations.}
\label{fig:sizeScalingTest}
\end{figure}
\end{center}

A direct test of the size-scaling approximation is provided in Fig.~\ref{fig:sizeScalingTest}, when survival probabilities obtained from a single pair of monomers embedded respectively in leaflets of 256 and 1024 lipids are displayed. The scaled survival curve $P_s(2,1024, t)^{4}$ matches  $P_s(2,256, t)$ at short times, while departing from it at longer times. This expected behavior  originates from trajectories showing a large separation between reacting pairs, allowed in the large system but forbidden (due to periodic boundary conditions) in the small system. It is also a signature of the size dependence of Brownian diffusion limited reaction dynamics in two-dimensional systems. Clearly, departure from the size scaling assumptions is visible on this curve, at large times. However, the most relevant region, as far as excimer formation is concerned, is the short time regime, and this is especially true at large probe concentrations $x$. Therefore, we consider that the size-scaling assumption is valid in our case. 

\begin{center}
\begin{figure}
\resizebox{0.42\textwidth}{!}{\includegraphics{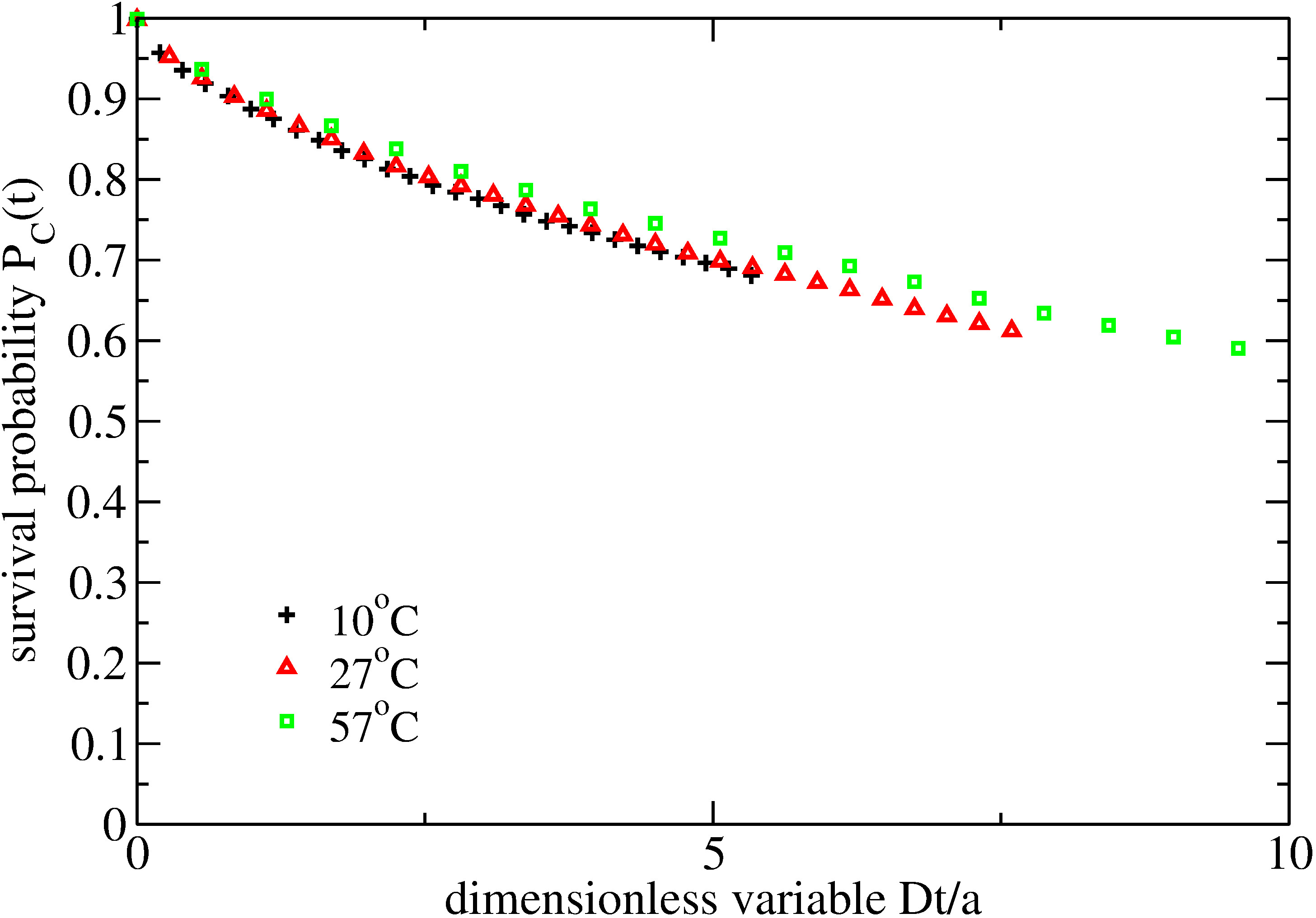}}
\caption{Time temperature superposition of the survival probability for DOPC bilayers. Simulated temperatures are 283, 300 and 330K. }
\label{fig:TimeTemperatureSuperposition}
\end{figure}
\end{center}

Fig.~\ref{fig:TimeTemperatureSuperposition} shows that survival probabilities obtained from trajectories at different temperatures almost superimpose, once a dimensionless time variable $u=Dt/a_0$ is used on the horizontal axis.


\subsection{Experimental titration curves}

The experimental data on monomer to excimer emission ratio are taken from Fig.~3 in reference~\cite{1990_Vauhkonen_Eisinger} (read from the graph) and displayed in Table~\ref{table:TitrationData}. The experimental values of the monomer fluorescence lifetime, which depends on temperature, are taken from Table~1 in reference~\cite{1990_Sassaroli_Eisinger}. The lifetimes of the isolated monomers are estimated from Fig.~5 in \cite{1990_Vauhkonen_Eisinger} to be respectively 140~ns for POPC at 20$^{\circ}$C and 160~ns for  DOPC at 10$^{\circ}$C. It must be stressed that the final results are very much dependent on the actual $\tau_M$ values.

\begin{center}
  \begin{table}
    {\scriptsize
  \begin{tabular}{||c|c|c||c|c|c||}
    \hline\hline
    & DOPC & & & POPC & \\
    $x$ & $1/J_M(x)$ & $J_M(x)$ & $x$ & $1/J_M(x)$ & $J_M(x)$ \\
    \hline
    0.003 & 1.02 & 0.98 & 0.003 & 1.03 & 0.97 \\
    \hline
    0.012 & 1.09 & 0.92 & 0.012	& 1.14 & 0.88 \\
    \hline
    0.034 & 1.31 & 0.76 & 0.033 & 1.52 & 0.66 \\
    \hline
    0.055 & 1.58 & 0.63 & 0.056 & 1.93 & 0.52 \\
    \hline
    0.102 & 2.45 & 0.41 & 0.102 & 3.25 & 0.31 \\
    \hline \hline
  \end{tabular}
}
  \caption{Monomer fraction $J_M(x)$ \textit{vs} probe concentration $x$}
\label{table:TitrationData}
\end{table}
\end{center}

\clearpage
\subsection{Fit of the excimer formation dynamics}

\begin{center}
  \begin{figure*}
    \begin{tabular}{cc}
      \resizebox{0.42\textwidth}{!}{\includegraphics{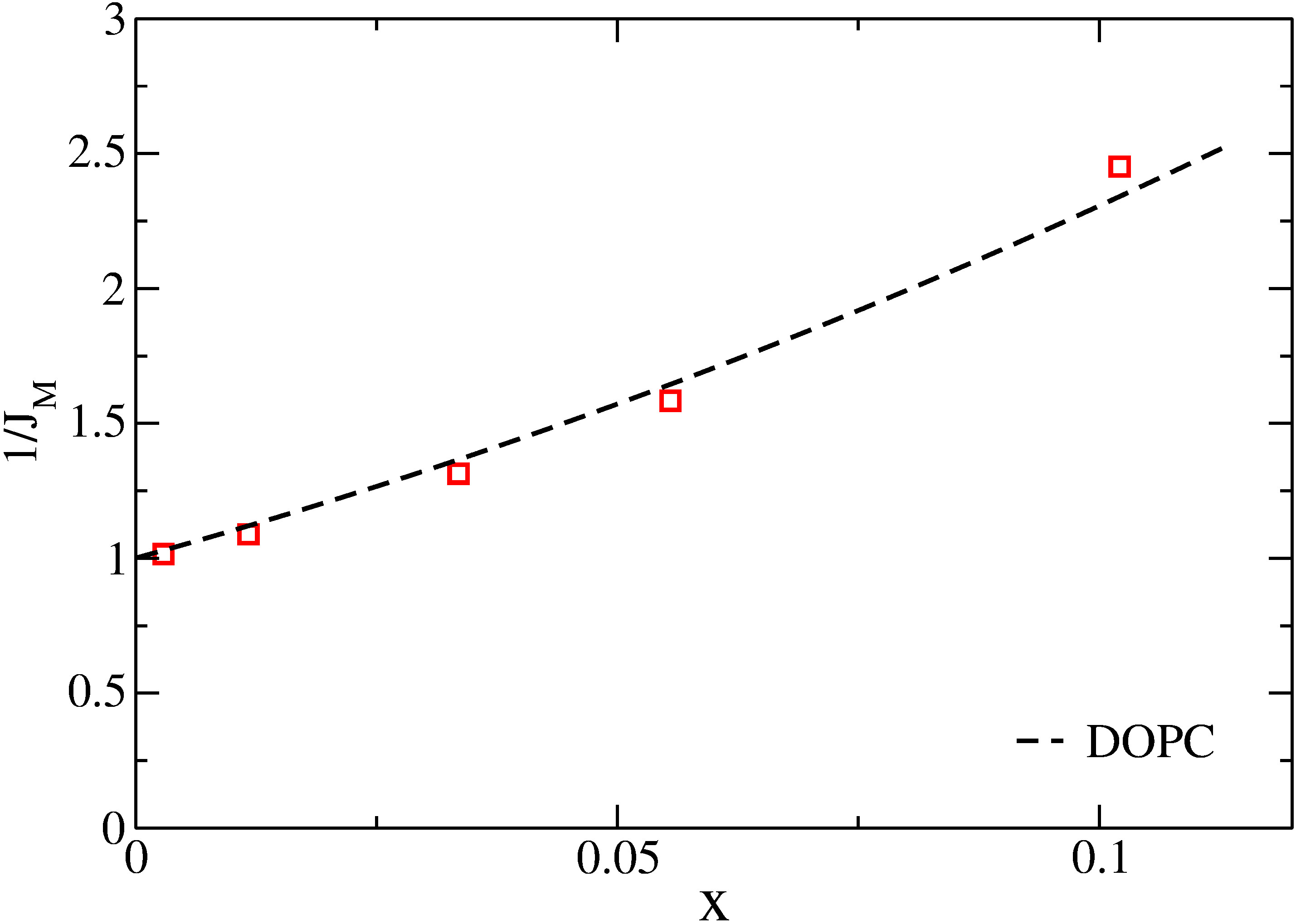}}
      &
      \resizebox{0.42\textwidth}{!}{\includegraphics{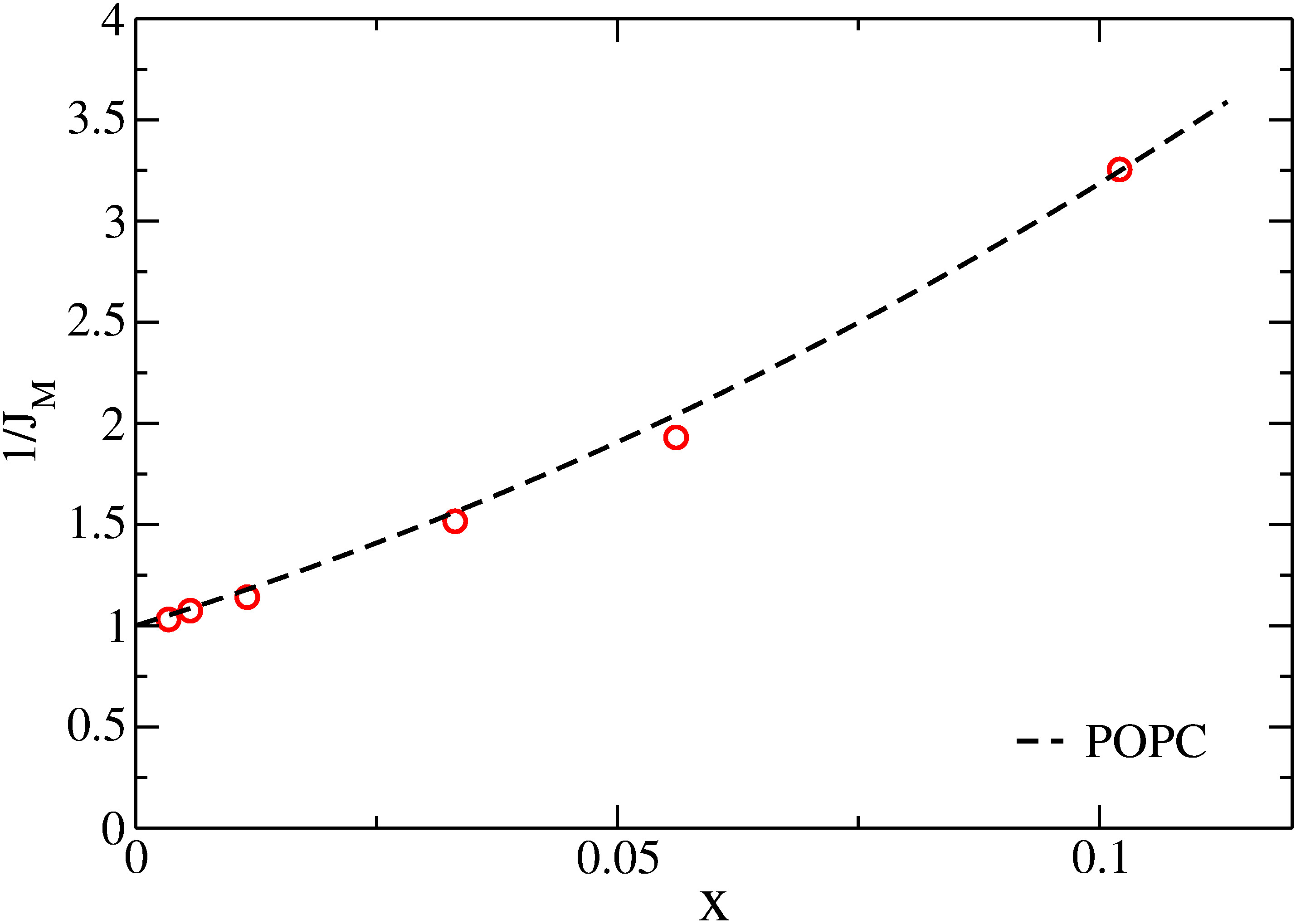}}\\
      \resizebox{0.42\textwidth}{!}{\includegraphics{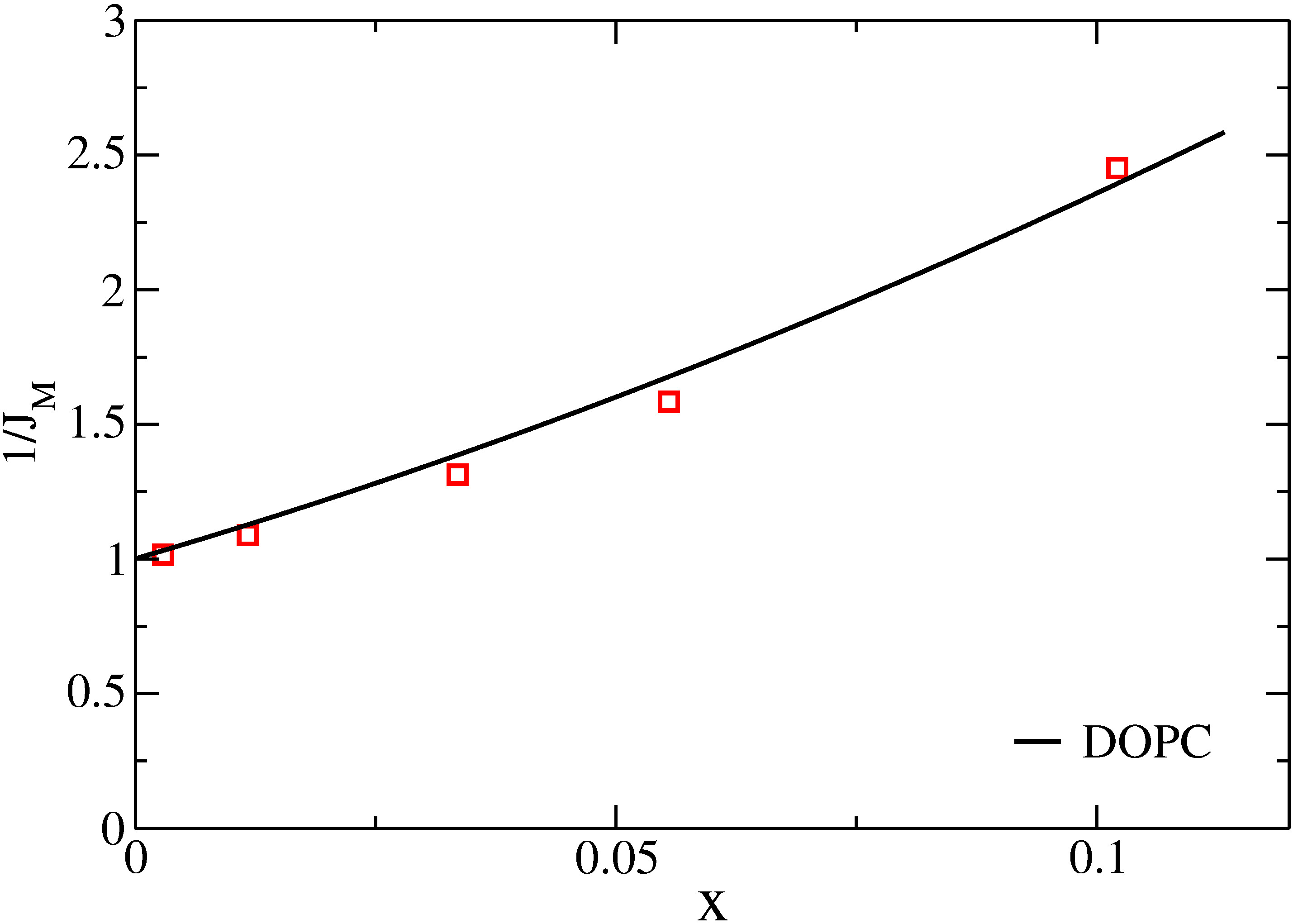}}
      &
      \resizebox{0.42\textwidth}{!}{\includegraphics{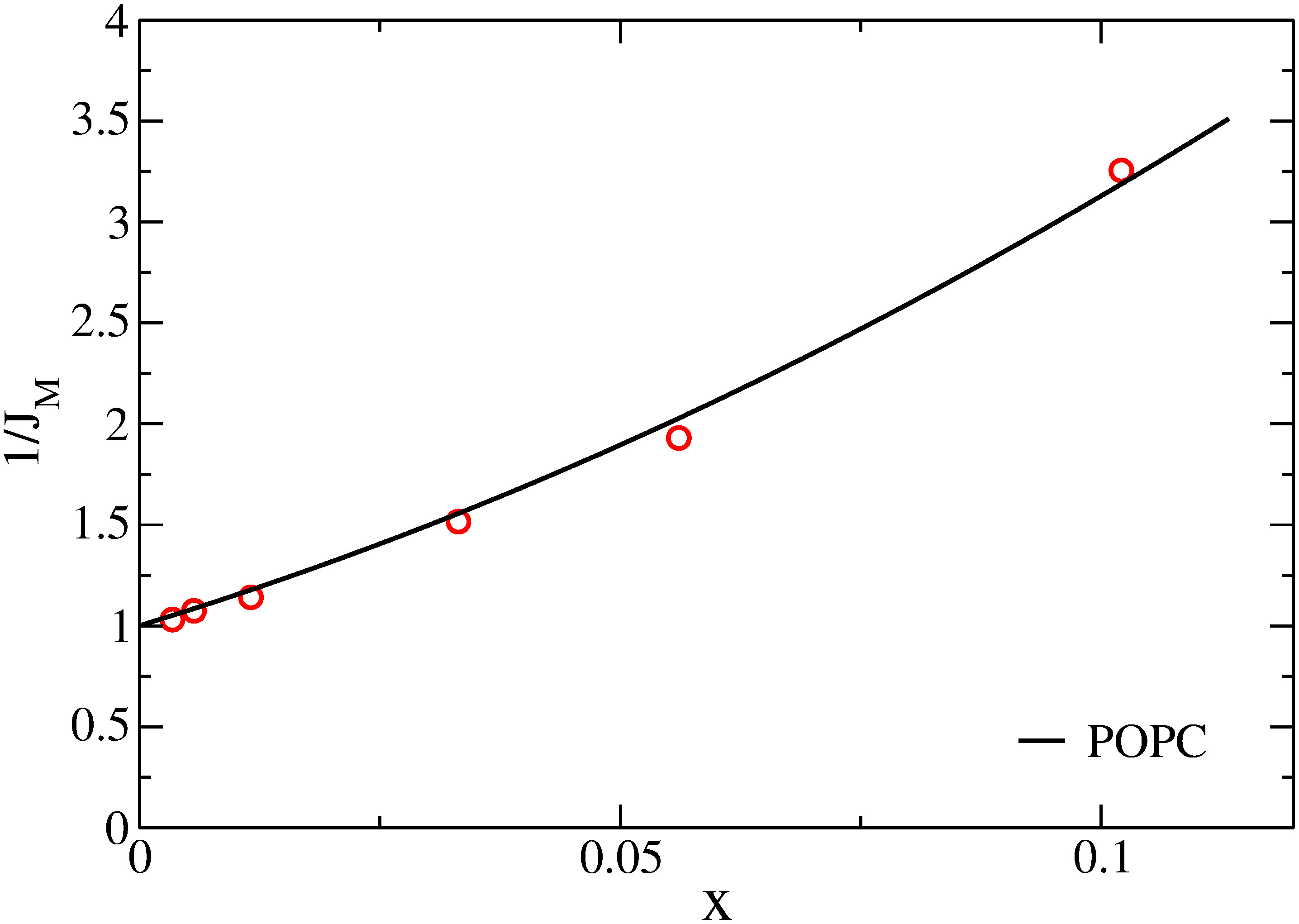}}
    \end{tabular}
    \caption{ \textit{Top left:} fit of $1/J_M(x)$ in the DOPC case when association is restricted to molecules within a single leaflet. \textit{Top right:} fit of $1/J_M(x)$ in the POPC case when association is restricted to molecules within a single leaflet. \textit{Bottom left:} fit of $1/J_M(x)$ in the DOPC case when association originates from both leaflets. \textit{Bottom right:} fit of $1/J_M(x)$ in the POPC case when association originates from both leaflets.}
\label{fig:fit-titration}
\end{figure*}
\end{center}


Fig.~\ref{fig:fit-titration}(\textit{top left}) shows the agreement between the experimental values and the numerical prediction extracted from the survival probability of Fig.~\ref{fig:logSurvival-DOPC}. The best agreement is obtained with ``numerical'' fluorescence lifetime $\tau_{M,CG}$ equals to 10.7~ns. The ratio between the experimental and the numerical lifetimes determines the acceleration factor $f = \tau_{M}/\tau_{M,CG}=15$, given that $\tau_{M}=160$~ns. The coarse-grained diffusion constant $D_{CG}$ is assumed to scale as $f D$, with $f$ the acceleration factor and $D$ the real time diffusion constant. This approach predicts a value $D=1.6~\mu\mathrm{m.s}^{-1}$ for DOPC molecules in a bilayer at 10$^{\circ}$C, if no interleaflet association are allowed.


When adjustment is made with POPC trajectories at 20~K, as shown in Fig.~\ref{fig:fit-titration}(\textit{top right}), the best fit is obtained with a numerical lifetime  $\tau_{M,CG}$ equal to 15.6~ns. The acceleration factor is now $f=9$ and the associated diffusion constant found for POPC at 20$^{\circ}$C is $D=4.0~\mu\mathrm{m.s}^{-1}$.


Fig.~\ref{fig:fit-titration}(\textit{bottom left}) shows the best agreement between the experimental  and numerical curves, when the survival probability $P_c(t)=P_s(t)\times P_o(t)$ corresponding to unrestricted  excimer formation is used. As the decay is now faster than in the previous case, a larger acceleration factor must be used to match the MD trajectories to the observed excimer formation rate. We obtain for DOPC at 10$^{\circ}$C a factor $\alpha=26$, leading to a diffusion constant $D = 1.0~\mu\mathrm{m.s}^{-1}$. A similar adjustment to the experimental data leads, for a POPC bilayer at 20$^{\circ}$C, to an acceleration factor $f=20$ and $D=1.8~\mu\mathrm{m.s}^{-1}$, shown in Fig.~\ref{fig:fit-titration}(\textit{bottom right}).


\section{Discussion}
\label{sec:Discussion}

We first observe that our procedure reproduce to some extend the upward curvature seen on the $1/J_M(x)$ plots, even though a larger curvature would be required to optimally fit the data. 

\begin{center}
\begin{table}
  \begin{tabular}{||c|c|c|c||}
    \hline\hline
    Lipid & Original $D$ & MD $D$ & MD $D$'  \\
    \hline
    POPC (293~K) & 11 & 1.8 & 4.0\\
    \hline
    DOPC (283~K) & 6 & 1.0 & 1.6 \\
    \hline\hline
     \end{tabular}
\caption{Comparison of diffusion constants, obtained with the milling crowd analysis of \protect\cite{1990_Vauhkonen_Eisinger} (column 2) and with the current approach, assuming reaction at first contact and allowing for interleaflet association (column 3) or excluding interleaflet association (column 4).}
\label{table:ComparisonDiffusion}
\end{table}
\end{center}

In Table~\ref{table:ComparisonDiffusion}, we compare our new values to those originally published by Vauhkonen et al. (Table~1 of ref.~\cite{1990_Vauhkonen_Eisinger}). Strictly speaking, the modeling of Vauhkonen et al excludes interleaflet excimer formation, and their values should be compared with our column~4. We find a significant reduction of the diffusion constant compared with their analysis.

Our diffusion constant in the case of allowed interleaflet formation is even smaller. On the one hand, if one trusts the coarse-grained MD trajectories, there is no reason to discard interleaflet association which occurs at a non negligible rate. In other words, $P_c(t)$ differs enough from $P_s(t)$ to alter significantly the final results. The pyrene probes by Vauhkonen et al. are the same, irrespective of the bilayers under consideration. Interleaflet association is expected to occur significantly for short lipids (\textit{e.g.} DMPC) but to be disfavored for longer lipid (\textit{e.g.} DOPC, DPPC, POPC) if there is a barrier preventing pyrene groups from interdigitating. Our coarse-grained model does not show evidence to this, but there is the possibility that it misses this point. United atoms simulations of pyrenyl groups grafted to saturated lipid chains in a DPPC fluid environment, as well as dipyrenyl groups grafted at the end of chains with variable length have been published~\cite{2006_Repakova_Vattulainen,2014_Franova_Ollila,2014_Franova_Vattulainen}. These studies suggests that interdigitation of the  pyrenyl group is possible if grafted to a 10~carbons length chain, as in the experiments considered here, and unlikely for shorter chains. Note that results obtained for saturated lipid chain may not be transposable to POPC and DOPC membranes. 

On the other hand, the results obtained by restricting pyrene association to same leaflet are closer from the body of published experimental coefficients. The Handbook of phospholipid bilayers published by Marsh~\cite{Marsh_HandbookLipidBilayers2} reports for DOPC at 25$^{\circ}$C the following two values: 6.3$\mu\mathrm{m}^2.s^{-1}$ (fluorescence correlation spectroscopy) and 1.8~$\mu\mathrm{m}^2.s^{-1}$ (electron spin resonance). Using an energy activation $E_a \sim 11~\mathrm{kJ.mol}^{-1}$, these values transpose to respectively 5 and 1.4~$\mu\mathrm{m}^2.s^{-1}$. It appears that the experimental diffusion coefficients reported in \cite{Marsh_HandbookLipidBilayers2} are widely scattered, and  much higher values of $D$ are even reported. For POPC at $20^{\circ}$C a FRAP value of 3.4~$\mu\mathrm{m}^2.s^{-1}$ is given. Again, it is consistent with our value in the absence of interleaflet association. Moreover, our new value is closer to FRAP than the original interpretation of the same data.

To conclude with diffusion coefficients, we find that when interleaflet association is allowed, our coefficients are a factor 2 smaller than those corresponding to fluorescent probe diffusion (FRAP or FCS). A better agreement is obtained by restricting pyrene association to probes occupying the same leaflet.

An outcome of the present work is the determination of the acceleration factor $f$ of a Martini CG lipid bilayer model.
At $10^{\circ}$C, $f$ lies between 15 and 25, depending on interleaflet association. At  $20^{\circ}$C it is reduced to 10-20. This is significantly higher than $f=4$ which is sometimes assumed. The $f=4$ factor corresponds indeed to the equivalent solvent (water) diffusion model, but seems higher as far as lipid motion is concerned. As the CG activation energy (14~kJ/mol) is likely to differ from the real activation energy, the agreement between CG model and reality should improve in the temperature range 300-330~K for which this model is supposed to perform best. Working at low temperatures enlarge the dynamical spread between numerical and real trajectories.

Coarse-graining raises the question of whether one can trust the resulting time correlation functions. There is no reason to assume that the acceleration factor applies uniformly for each time interval. If one denotes $C_{AB}^{CG}(t)=\langle A(t)B(0)\rangle$ the correlation function of a pair of observables, and  $C_{AB}(t)=\langle A(t)B(0)\rangle$ the experimental (or all-atom simulated) counterpart, a correspondence $C_{AB}^{CG}(T(t)) = C_{AB}(t)$ is expected, with possibly a non-linear monotonous time correspondence $T(t)$ between the two correlations.

An homogeneous acceleration factor corresponds to $T(t)=t/f$. An inhomogeneous correspondence could explain why is the short time collisional dynamics faster than the long time diffusion process. Our work sets bounds on such a phenomenon. If interleaflet association is forbidden, the agreement between numerical and experimental values of $D$ is consistent with an homogeneous acceleration factor. If interleaflet association is allowed,  the time-scale dependent effective factor $f=25$ (DOPC) or $f=20$ (POPC) that acts on short time separation,  reduced by a factor 2, would be again consistent with a diffusion coefficient of respectively 2 and 3.6~$\mu\mathrm{m}^2.s^{-1}$.

Another way of slowing down the collision dynamics would be to challenge the diffusion limited character of the excimer formation, by requiring repeated collisions between monomers prior to complexation. This effect can be introduced by changing the excimer formation rate $q$ in equation~(\ref{eq:SurvivalFunctionSampling}). For DOPC, interleaflet association allowed, a diffusion coefficient $D=2.2~\mu\mathrm{m}^2.s^{-1}$ can be obtained by reducing $q$ to a value such that it would take 1.1~ns of real time for the excimer to form, with both monomers maintained at close vicinity. This extra time could be related to a need of the terminal subgroups to align parallel, prior to excimer formation. 

A drawback of slowing back in such a way the dynamics is that the effective survival probability becomes closer and closer to exponential shape, and therefore does not lead to the desired upward curvature of the $1/J_M(x)$ plot. Therefore, we do not find that it improves the agreement with experimental data to introduce such an intrinsic excimer formation time. A clue to whether such a delay is necessary, and its order of magnitude could be obtained from time-resolved excimer formation.

\section{Conclusion}

We have presented a very general and efficient method to derive association kinetic coefficients in the case of diffusion limited processes, based on realistic molecular motions in lipid bilayers. The time-dependent reaction rate is derived from a survival probability, which itself comes from \textit{a posteriori}  trajectory analysis. A virtual relabeling of the simulated molecules enables us to efficiently gather collision statistics. No assumptions are made regarding the Brownian nature of the molecular motion, and intramolecular fluctuation effects are taken into account, within the limits of the coarse-graining procedure. 

Our main assumptions are that lipid probes behave similarly to the major lipid components, and do not interact strongly with each other. We also require the dynamical acceleration factor of the coarse-grained numerical model to be uniform over the relevant time scales.

It is first shown that simulations at different sizes and temperatures can be compared, and used for the purpose of studying the collision dynamics between virtually labeled lipids. A likely capture radius $\rho_c=0.5~nm$ is obtained from a pair correlation analysis of the reactive groups.

Two main situations are considered, depending on whether interleaflet excimer association is allowed or not. Based on CG simulations considerations, interleaflet association should be allowed, but lead to small values of the diffusion coefficients. If one forbids interleaflet association, given that there could barriers opposing it that our CG model misses, the resulting diffusion constants fall in a range comparable to other independent experimental procedures. Between these extremes, there is also the possibility that interleaflet excimer formation occurs, but comparatively less frequently than intraleaflet excimers.

In any case, our approach leads to a significant reduction of the diffusion coefficients compared to those originally published. We suggest that the coefficients found in similar diffusion limited processes,  which often lie above the values obtained from long range fluorescent probes diffusion, should be reanalyzed along the lines of our present approach. We suspect that these diffusion coefficients are overestimated due to analysis bias, as our example shows.

Our standing is that we have possibly explained the discrepancy between these two different approaches to lipid diffusion determination. More modeling work is needed in order to determine the degree of interdigitation of pyrene groups pertaining to different leaflets, as well as regarding the effect of non trivial dispersion forces between resonant fluorescent groups.

The application of this technique to complex membranes with dynamical heterogeneities would certainly be interesting, now 
that a convenient framework for diffusion limited reaction rate analysis, based on molecular dynamics, is available. 

\section*{Author contributions}
P.A. did the simulation and numerical work, the data fitting and composed the figures (4-9). F.T. wrote the manuscript. Both authors contributed to the theoretical model and to figures (1-3).

\section*{Acknowledgments}

F.T. would like to thank G.~Duportail, C.M.~Marques and J.~L\'{e}onard for discussions. The authors acknowledge support from the High Performance Cluster Equip@Meso of the University of Strasbourg for providing computational resources.


\end{document}


\title{Supplementary Material (Excimer formation of pyrene labeled lipids interpreted by means of coarse-grained molecular dynamics simulations)}   
\author{P. Ayoub, F. Thalmann$^{\ast}$}
\address{Institut Charles Sadron, Université de Strasbourg, CNRS UPR~22, 23 rue du Loess, Strasbourg Cedex, F-67037, France.\\
$^{\ast }$fabrice.thalmann@ics-cnrs.unistra.fr
}
\maketitle

\markboth{Ayoub et al.}{Supplementary Material}

\appendix
\section{Equivalence between exponential survival rate and rational distribution}

Let $x=x^{\ddagger}$ be the value that makes the excimer formation probability equal to $1/2$: 
\beq
J_M(x^{\ddagger}) = J_E(x^{\ddagger}) = 1/2. 
\eeq
Let us assume an exponential survival probability $\Pg(t)=\exp(-t/\tau_c(x))$. From eq.~(6) comes
\beq
\frac{1}{J_M(x)}= 1 + \frac{\tau_M}{\tau_c(x)}.
\eeq
$x^{\ddagger}$ is the concentration for which the collision time $\tau_c(x^{\ddagger})$ equals the monomer spontaneous decay time $\tau_M$.

The independent pair assumption, combined with an exponential survival probability reads
\beq
\Pg(t)= \exp\left(-\frac{t}{\tau_c(x^{\ddagger})}\frac{x}{x^{\ddagger}}\right)
\label{eq:from_exponential_rate}
\eeq
and
\beq
\frac{1}{J_M(x)}= 1 + \frac{\tau_M}{\tau_c(x^{\ddagger})}\frac{x}{x^{\ddagger}}=1 + \frac{x}{x^{\ddagger}}
\eeq
With monomer in excess, the situation corresponds to an ideal pseudo-first order reaction kinetics, consistent with an exponential decay of the excited monomer population.  The decrease rate of the excited probes reads $\dd [M^{\star}] = -(K(x)+\tau_M^{-1})[M^{\star}]\dd t$, leading to the monomer emission intensity
\beq
\frac{1}{J_M(x)} = 1+\tau_M K(x).
\eeq
In a mean-field, constant rate bimolecular mechanism, $K(x)$ is proportional to $x$ and therefore again
\beq
\frac{1}{J_M(x)} = 1+\frac{x}{x^{\ddagger}}.
\label{eq:from_constant_rate}
\eeq
It is clear from eqs.~(\ref{eq:from_exponential_rate}) and (\ref{eq:from_constant_rate}) that both approaches coincide. 

\subsection{Multiple passage times}

The possible strategy, which we did not consider in our main paper,  consists in introducing a generalization $\Pg^{(n)}$ of the survival probability $\Pg$. Let $\Pg^{(n)}$ be the probability that the excited monomer has suffered exactly $n$ collisions during the excitation time $t$. The first term of the sequence $\Pg^{(0)}$ coincides with the survival probability $\Pg$. The time derivative $-\dd \Pg^{(n)}/\dd t $ gives the distribution of $n^{th}$ collision time. By definition 
\beq
\sum_{n=0}^{\infty} \Pg^{(n)}(t) = 1,
\eeq
and in the case of independent collision events with constant rate, one recovers a Poisson distribution
\beq
P_{\mathrm{Poisson}}^{(n)}(t) = \frac{ t^n}{n! \tau_c^n}\exp\left(-\frac{t}{\tau_c}\right).
\eeq

If trajectories and excimer formation reaction are uncorrelated, one can introduce an effective survival probability 
\beq
P_{\mathrm{eff}}(t) = (1-W_E) \sum_{n=0}^{\infty} W_E^n \Pg^{(n)}(t),
\eeq
with $W_E$ the probability to form an excimer at a given encounter, such that  eq.~(13) applies with $P_{\mathrm{eff}}(t)$ substituted for $\Pg$. 

When using this strategy in the Poissonian case, one finds that the effective survival probability remains exponential with an effective decay time $(1-W_E)\tau_c$. As a byproduct, the titration curve is strictly independent of $W_E$. Therefore \textit{it is not possible to determine $W_E$ in the case of an uncorrelated, Poissonian collision dynamics process}. Irrespective of $W_E$ the inverse monomer probability reads $1/J_M(x)=1+x/x^{\ddagger}$. Some deviation from pure exponential behavior of the survival function is necessary if one wants to observe and fit a nonlinear curve $1/J_M(x)$. 

\section{Dynamical quantities}
\label{sec:DynamicalQuantities}

The Martini CG-DOPC consists of 14 beads, out of which chain terminal subgroups correspond to beads 7 to 9 on the one hand, and 12 to 14 on the other hand. Let us denote by $\vec{r}_{i,\alpha}(t)=(x_{i,\alpha}(t), y_{i,\alpha}(t), z_{i,\alpha}(t))$ the trajectory of a bead, $i$ being the index of the molecule and $1<\alpha<14$ the index of the bead. Center of mass (CM) and terminal subgroup (TS) positions are defined as 
\beqn
\vec{r}_{i,CM} &=& \frac{1}{14} \sum_{\alpha=1}^{14} \vec{r}_{i,\alpha}(t)\\
\vec{r}_{i,TS2} &=& \frac{1}{3} \sum_{\alpha=12}^{14} \vec{r}_{i,\alpha}(t)
\eeqn

Mean square displacements are obtained from unwrapped trajectories, using \textit{e.g.} the \textit{g\_rdf} Gromacs analysis tool.  
\beq
g_{TS2}(t) = \frac{1}{N_t} \mean{[x_{i,TS2}(t)-x_{i,TS2}(0)]^2}
\eeq

Collisions occurs when two terminal subgroups are found within an interaction sphere of radius $\rho_c$. In other words, two groups are within reaction distance if 
\beq
d_{MID}(\vec{r}_{i,TS2}(t), \vec{r}_{i,TS2}(t))^2 \leq \rho_c^2
\eeq
where $d_{MID}$ stands for the Euclidean distance between groups $i$ and $j$ in the minimal image distance convention. 

\subsection{Simulation details}

Simulations were performed with Gromacs 4.6 on the \textit{High Performance Cluster} of the University of Strasbourg. We have simulated symetric patches of bilayers made of respectively 512 DOPC and 512POPC \textit{Martini} coarse-grained molecules. Constant volume simulations (for a pressure close to 1~bar) and Nose-Hoover thermostat were used. Note that constant pressure simulations would have given indentical results. Four trajectories of $5\times 10^{6}$ steps were used for each lipid (DOPC at 283K, POPC at 293K). 

Trajectory frames are collected every 200~steps (every 6~ps) for the collision analysis. This frame frequency is significantly higher than what is normally required  when computing membrane equilibrium properties. The simulation time was found to be sufficient, given the spontaneous monomer deexcitation dynamics which restricts excimer formation to the initial part of $\Pg(t)$. 

A patch of 2048 DOPC lipids was also simulated to provide the size-scaling assumption comparison in Figure~7.
%
\subsection{Numerical integration of eq.(6)}

Numerical sampling of the survival functions, such as outlined in eq.~(13), provides a function $P_{\mathrm{num}}(t)$ sampled at equal times up to a cut-off value $t_{co}$. The behavior at larger times is, within good approximation, exponential. 

The long times behavior does not dominate the quadrature in eq~(6) for two reasons. First, the integration is naturally cut-off by the spontaneous monomer  decay rate $\exp(-t/\tau_M)$. Second, increasing $x$ has the effect of putting more weight on the short time regimes, making the long time regime irrelevant as far as computing $J_M$ and $J_E$ is concerned.

A trapezoid integration rule was found to give accurate enough results. Such an integration is required for each value of the acceleration factor, and each choice for $x$. The acceleration factor $f$ was determined by trial and errors, until best visual fit to the titration curves in Figs~9,10,11,10.

\bibliographystyle{plain}
\bibliography{/home/fabrice/Documents/TEX/Bibtex/ics,/home/fabrice/Documents/TEX/Bibtex/Books}